\documentclass{PoS}

\usepackage{amsfonts}
\usepackage{amsmath}
\usepackage{subfigure} 

\graphicspath{{./figs/}}

\title{
  Recent progress in calculation of $B_K$ using staggered fermions
}

\ShortTitle{
  Recent progress in staggered $B_K$
}

\author{\speaker{Weonjong Lee}, 
  Yong-Chull Jang, Hwancheol Jeong, Jangho Kim,
  Kwangwoo Kim, Seonghee Kim, Jaehoon Leem, Boram Yoon \\
  Lattice Gauge Theory Research Center, CTP, and FPRD, \\
  Department of Physics and Astronomy,
  Seoul National University, Seoul, 151-747, South Korea \\
  E-mail: \email{wlee@snu.ac.kr}}

\author{Taegil Bae \\
  Korea Institute of Science and Technology Information, 
  Daejeon, 305-806, South Korea \\
  E-mail: \email{esrevinu@gmail.com}}

\author{Chulwoo Jung, Hyung-Jin Kim \\
  Physics Department, Brookhaven National Laboratory,
  Upton, NY11973, USA \\
  E-mail: \email{chulwoo@bnl.gov}}

\author{Jongjeong Kim \\  
  Physics Department,
  University of Arizona,
  Tucson, AZ 85721, USA \\
  E-mail: \email{rvanguard@gmail.com}}

\author{Stephen R. Sharpe\\
  Physics Department, University of Washington, 
  Seattle, WA 98195-1560, USA \\
  E-mail: \email{sharpe@phys.washington.edu}}

\author{SWME Collaboration}

\abstract{ We report on recent progress in the calculation of $B_K$
  using HYP-smeared improved staggered fermions on the MILC asqtad lattices.
  We have added measurements on fine ($a\sim 0.09\;$fm)
  and superfine ($a\sim 0.06\;$fm) ensembles
  at different values of the light sea quark mass ($a m_\ell$),
  as well as increased the statistics on some other ensembles.
  We find that the results on the fine lattices show
  a significantly stronger $am_\ell$ dependence than those on the superfine
  and coarse ($a\sim 0.12\;$fm) lattices. 
  We discuss different methods for accounting for these new results
  when doing the $a m_\ell$ and continuum extrapolations.}

\FullConference{
  The 30 International Symposium on Lattice Field Theory - Lattice 2012,\\
  June 24-29, 2012\\
  Cairns, Australia
}

\begin{document}

\section{Introduction} 

The calculation of the kaon mixing matrix element $B_K$ is
one of the successes of lattice QCD. Results with all errors
controlled are available with several 
fermion discretizations~\cite{FLAG,LatticeAverages}.
The value for $B_K$ is a key input into 
the standard model prediction for $\epsilon_K$.
At present there is an $\approx 3\sigma$ tension 
between this prediction and the experimental value 
(if one uses the ``exclusive'' $V_{cb}$ obtained using lattice 
QCD)~\cite{laiho-2010-1,lunghi-2011-1,swme-jyc}.
Thus it is important to continue to further improve
the lattice calculations.

In this proceedings we provide an update on our results for $B_K$.  
These are
obtained using improved staggered fermions, specifically HYP-smeared
valence quarks on asqtad sea-quarks.  We describe here results
obtained using chiral extrapolations
based on SU(2) staggered chiral perturbation theory (SChPT),
which is our most reliable procedure~\cite{swme-1,swme-2}.  Details of
the fitting functions and the analysis method are given in 
Ref.~\cite{swme-1,swme-2}, and are not repeated here.  

Table \ref{tab:milc-lat} lists all the ensembles on which
we have calculated $B_K$, and notes which results have changed
in the last year. In particular, since Lattice 2011 we have
accumulated much higher statistics on the F2 and S1 ensembles 
(in addition to a small increase on the U1 ensemble)
and added new measurements on ensembles F3, S2 and S3.
Of these, the most important updates are those on 
ensembles F2, F3, S2, and S3,
since they give information on the light sea quark mass dependence
which was previously lacking on the fine and superfine ensembles.
We focus on this dependence here.
\begin{table}[h!]
\begin{center}
\begin{tabular}{ c | c | c | c | c | c }
\hline
$a$ (fm) & $am_l/am_s$ & geometry & ID & ens $\times$ meas 
& status \\
\hline
0.12 & 0.03/0.05  & $20^3 \times 64$ & C1 & $564 \times 9$ & old \\
0.12 & 0.02/0.05  & $20^3 \times 64$ & C2 & $486 \times 9$ & old \\
0.12 & 0.01/0.05  & $20^3 \times 64$ & C3 & $671 \times 9$ & old \\
0.12 & 0.01/0.05  & $28^3 \times 64$ & C3-2 & $275 \times 8$ & old \\
0.12 & 0.007/0.05 & $20^3 \times 64$ & C4 & $651 \times 10$ & old \\
0.12 & 0.005/0.05 & $24^3 \times 64$ & C5 & $509 \times 9$ &  old \\
\hline
0.09 & 0.0093/0.031 & $28^3 \times 96$ & F3 & $949 \times 9$ & \texttt{new} \\
0.09 & 0.0062/0.031 & $28^3 \times 96$ & F1 & $995 \times 9$ & old \\
0.09 & 0.0031/0.031 & $40^3 \times 96$ & F2 & $959 \times 9$ & \texttt{update} \\
\hline
0.06 & 0.0072/0.018 & $48^3 \times 144$ & S3 & $593 \times 9$ & \texttt{new} \\
0.06 & 0.0036/0.018 & $48^3 \times 144$ & S1 & $749 \times 9$ & \texttt{update} \\
0.06 & 0.0025/0.018 & $56^3 \times 144$ & S2 & $799 \times 9$ & \texttt{new} \\
\hline
0.045 & 0.0028/0.014 & $64^3 \times 192$ & U1 & $747 \times 1$ & \texttt{update} \\
\hline
\end{tabular}
\end{center}
\caption{MILC asqtad ensembles used to calculate $B_K$. 
$a m_\ell$ and $a m_s$ are the masses, in lattice units,
of the light and strange sea quarks, respectively. ``ens'' indicates
the number of configurations on which ``meas'' measurements are made.
Note that the numbering of
the ID tags on the fine and superfine lattices do not follow the ordering
of $a m_\ell$.
}
\label{tab:milc-lat}
\end{table}

\section{Chiral fits}
In our numerical study, our lattice kaons are composed of valence
(anti)quarks with masses $m_x$ and $m_y$. These are, respectively, the
masses of the valence $d$ and $s$ quarks.
On each MILC ensemble, we use 10 valence masses:
\begin{equation}
  am_x, am_y = am_s \times {n}/{10} \qquad \text{with} \qquad 
  n = 1,2,3,\ldots,10,
\end{equation}
where $am_s$ is the nominal strange sea quark mass.
In our standard fits we extrapolate to $am_d^{\rm phys}$
using the lowest 4 values for $am_x$ (the ``X-fit''---done
at fixed $am_y$),
and then extrapolate to $m^\textrm{phys}_s$ using the
highest 3 values of $am_y$ (``Y-fit'').
As described in Ref.~\cite{swme-1}, these choices keep us
in the regime where we expect next-to-leading order (NLO)
SU(2) ChPT to be reasonably accurate.
The X-fits described here are done to the form predicted by
NLO partially quenched SChPT (which is
given in Refs.~\cite{swme-1,ref:VdWS}), 
augmented by analytic and generic non-analytic
terms of next-to-next-to-leading order (NNLO)
and a single analytic term of next-to-next-to-next-to-leading order (NNNLO).
We use the Bayesian method to constrain the higher order coefficients,
as described in Refs.~\cite{swme-1,swme-2,wlee-2011-2}.
The $am_y$ dependence (which is not controlled by ChPT) is very close
to linear and we use a linear fit for our central values.  We dub this
entire fitting procedure the ``4X3Y-NNNLO fit'', and use it for our
central values. 


%
%
%
\begin{figure}[t!]
\centering
\subfigure[F3]{\includegraphics[width=0.49\textwidth]
  {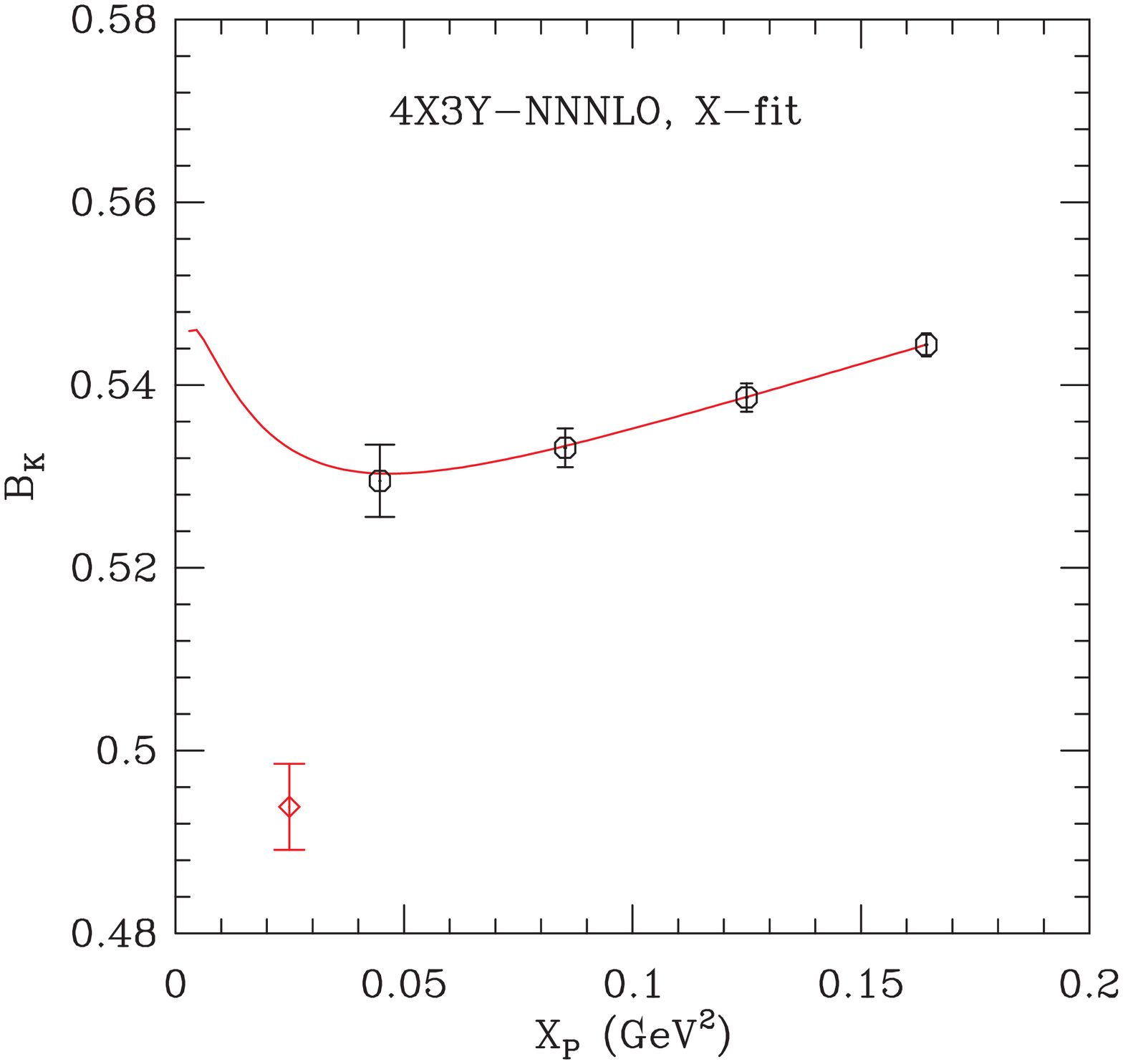}}
\subfigure[F2]{\includegraphics[width=0.49\textwidth]
  {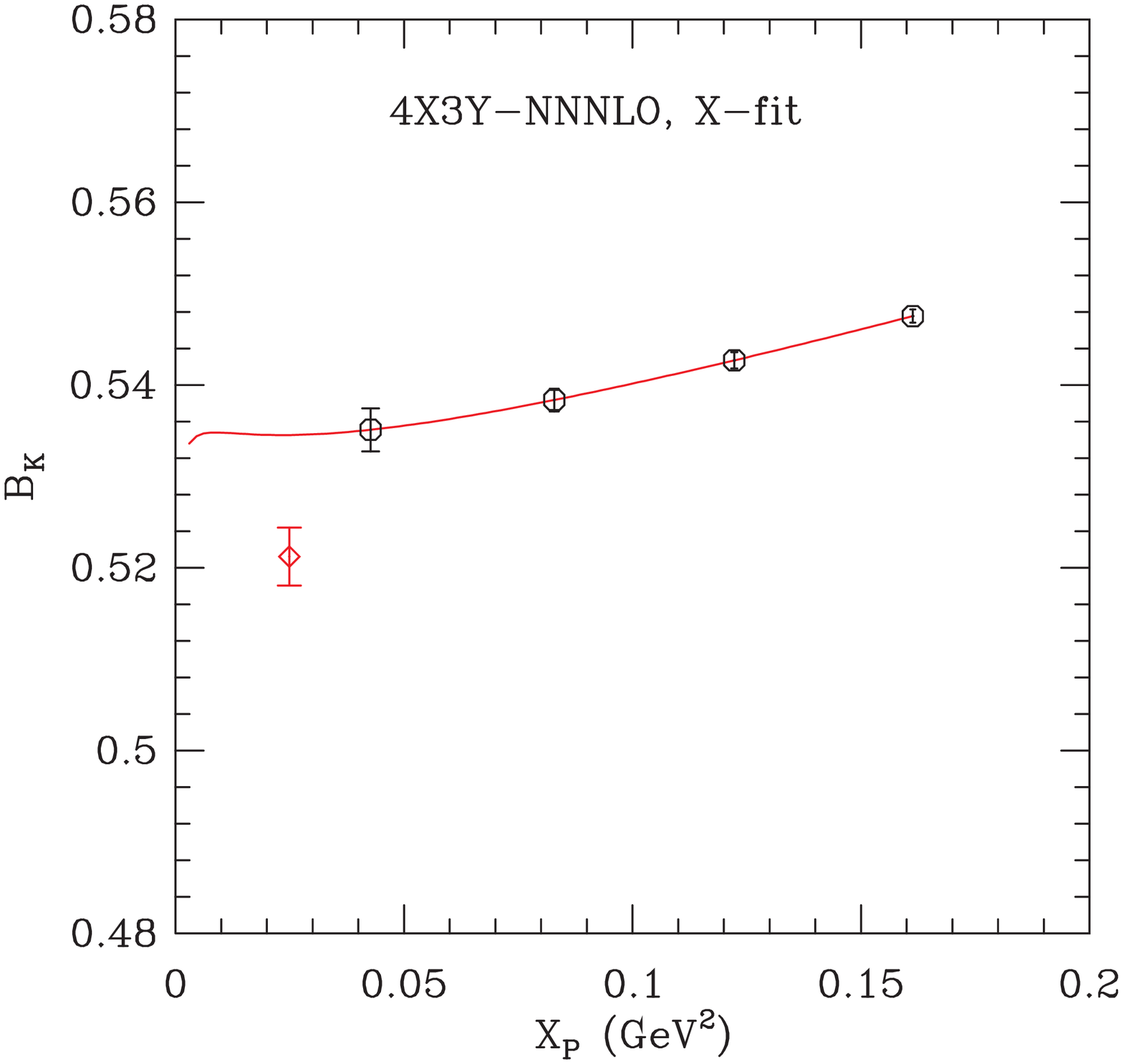}}
\caption{X-fits to $B_K(1/a)$ for the F3 and F2 ensembles.
The red diamond is explained in the text.}
\label{fig:su2-4x3y-nnlo:F3+F2}
\end{figure}
\begin{figure}[bt!]
\centering
\subfigure[S3]{\includegraphics[width=0.49\textwidth]
  {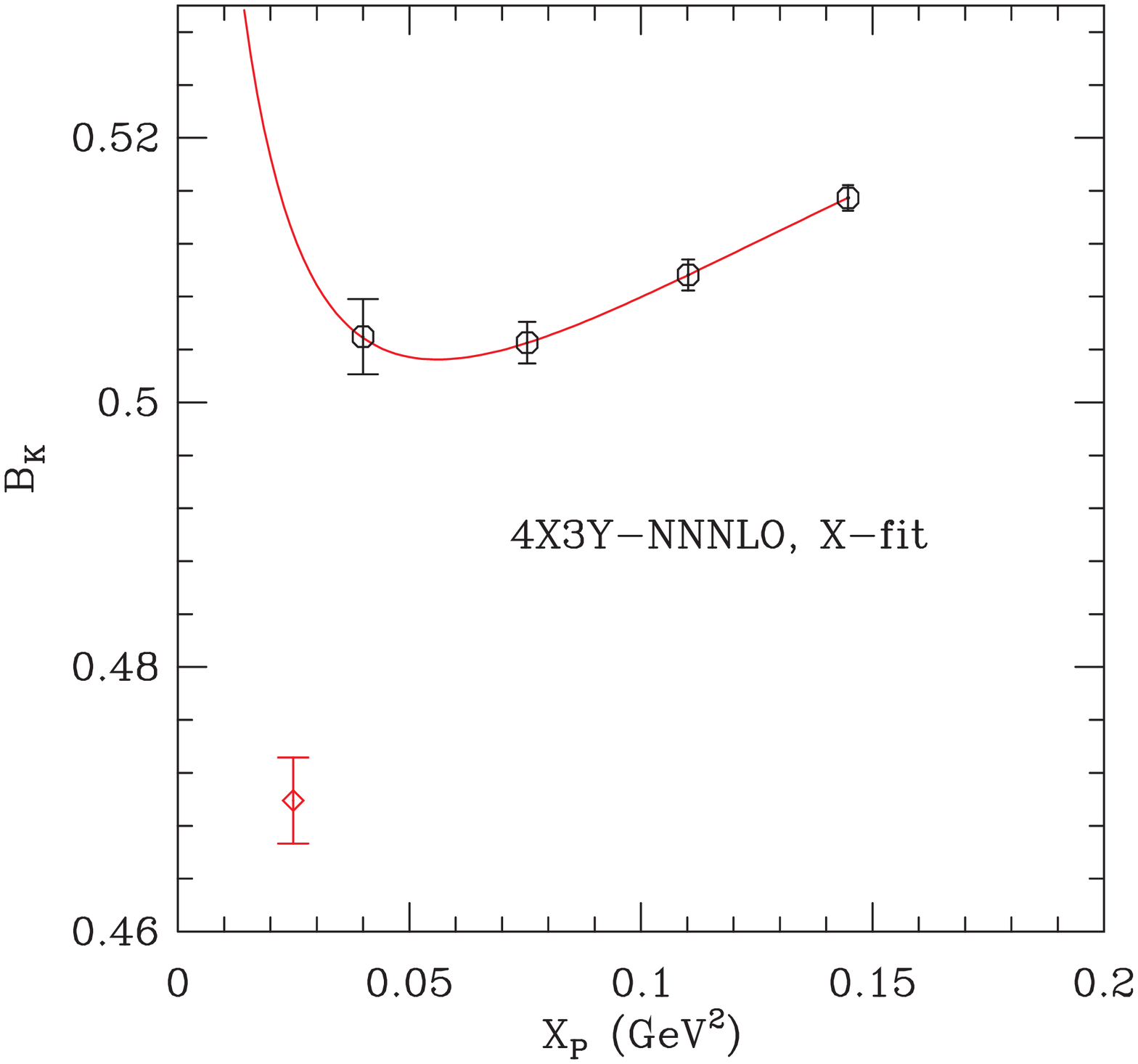}}
\subfigure[S2]{\includegraphics[width=0.49\textwidth]
  {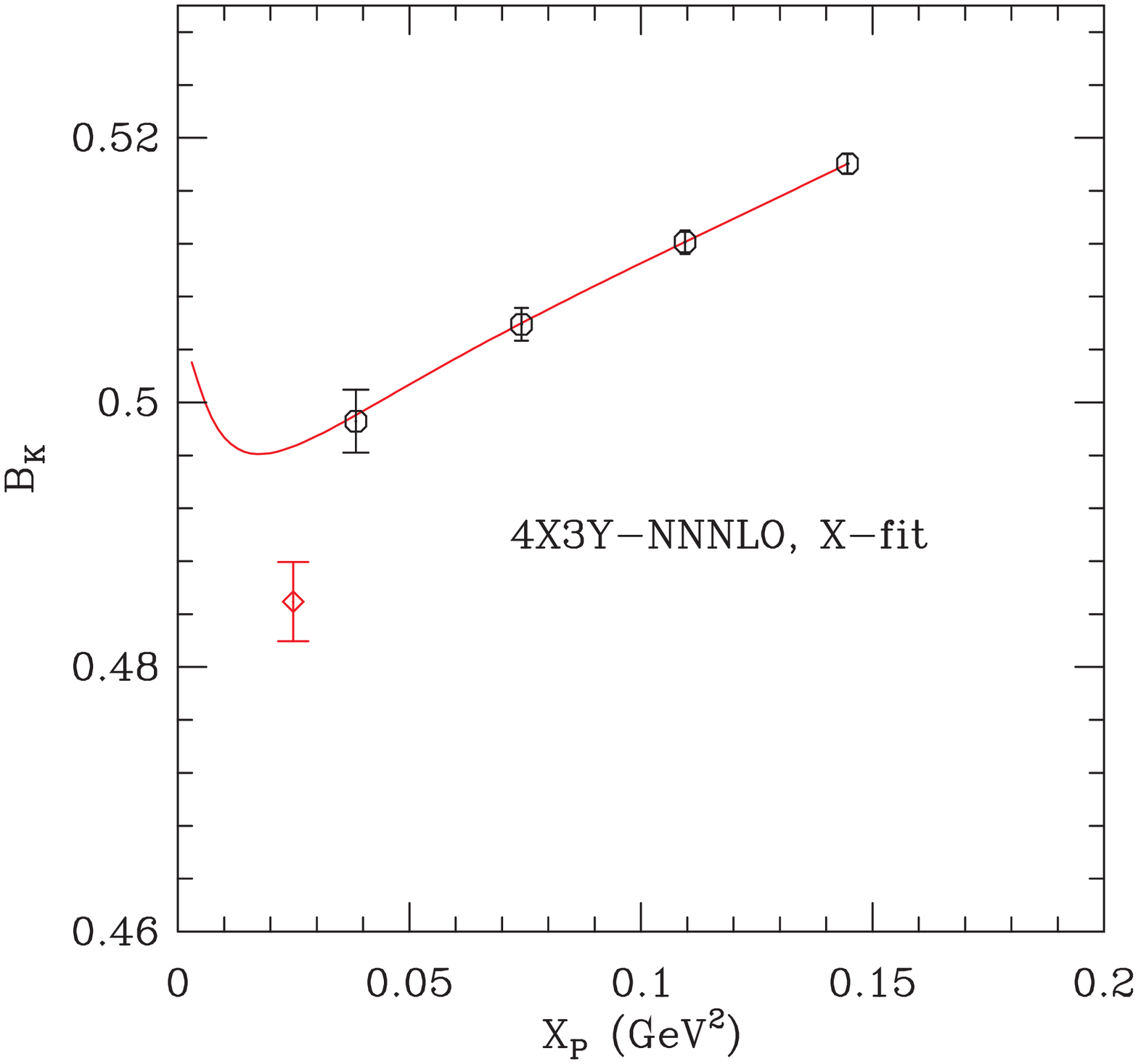}}
\caption{X-fits to $B_K(1/a)$ for the S3 and S2 ensembles. Notation
as in Fig.~\protect\ref{fig:su2-4x3y-nnlo:F3+F2}.}
\label{fig:su2-4x3y-nnlo:S3+S2}
\end{figure}
%
%
The X-fits on the new fine ensembles
are shown in Fig.~\ref{fig:su2-4x3y-nnlo:F3+F2},
while those on the new superfine ensembles are shown in
Fig.~\ref{fig:su2-4x3y-nnlo:S3+S2}.
In each panel, a fit to the SU(2) SChPT form is shown,
with $X_P$ the squared-mass of the Goldstone-taste valence
$\bar x x$ pion, whose mass is very close to linear to $ a m_x$.
Also shown (as the red diamond) is the result obtained after
(a) extrapolating $a m_x\to a m_d^{\rm phys}$, 
(b) removing the known taste-breaking in the pion masses appearing in the
chiral logarithms, and
(c) setting the light sea-quark
mass to its physical value in the chiral logarithmic terms.
See Ref.~\cite{swme-1} for details of this procedure. 
We have incorporated finite volume corrections,
as predicted by NLO SChPT,
into the fitting function~\cite{wlee-2011-3}.

A notable feature of both figures is that the difference between
the extrapolated/corrected result and the fit curve is larger on
the ensembles with larger $a m_\ell$ (F3 and S3). This is due
almost entirely to the larger shift required to
bring the sea-quark pion mass to its physical value
on these lattices (correction (c) above).
The other aspects of the extrapolation/correction are similar for
the pairs of lattices.
We return to this point below.

\section{Dependence On Light Sea Quark Mass}
\label{sec:am_l}
\begin{figure}[t!]
\centering
\includegraphics[width=0.6\textwidth]{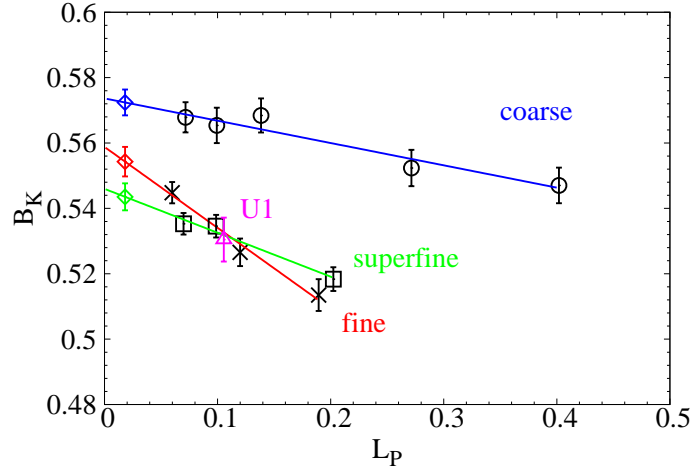}
\caption{$B_K(\mu=2\,\text{GeV})$ vs. $L_P \, (\text{GeV}^2)$,
where $L_P$ is a squared mass of the Goldstone-taste
pion composed of two light sea quarks
($\bar{\ell}\ell$). 
}
\label{fig:bk-su2-lp:all}
\end{figure}
In Fig.~\ref{fig:bk-su2-lp:all}, we show the dependence of $B_K$
on the light sea quark mass for all lattice spacings,
including the U1 point for completeness.
We use $L_P$ as a proxy for $a m_\ell$, since these quantities
are proportional to very good accuracy.
We have run $B_K$
to a common renormalization scale $\mu = 2 \,\text{GeV}$ 
(in the $\overline{\rm MS}$ scheme), so that the results are
directly comparable.
We expect, at NLO, only a linear behavior on $L_P$.

As can be seen from the plot, a linear fit works well for each
of the three lattice spacings for which we have multiple values
of $am_\ell$.
The results on the coarse lattices, 
which were presented in Ref.~\cite{swme-2},
show a mild dependence on $L_P$. The slope, $-1/(2.9\;{\rm GeV})^2$,
is consistent with naive dimensional analysis, which predicts
a magnitude of ${\cal O}(1/\Lambda_\chi^2)$ with $\Lambda_\chi\approx 1\;$GeV.
The slope on the superfine lattices, which is a new result,
has a slightly larger magnitude than that on the coarse lattices 
but is comparable.
What is striking, however, 
is the slope on the fine lattices, whose magnitude is about twice that
on the coarse lattices (slope $\approx -1/(2\;{\rm GeV})^2$).
Its value per se is reasonable---what is unexpected is that it should
differ so much from that at the other lattice spacings.
Since we are in the regime where corrections are small ($\le 10\%$)
in both chiral and continuum extrapolations, one would expect that
the slope would depend only weakly on $a^2$, and any dependence
should be linear. Instead, our results appear to require a contribution
to $B_K$ proportional to $am_\ell (a^2)^2$, since only with a quadratic
dependence on $a^2$ can the magnitude of the slope increase and then decrease
as $a^2$ is raised.

This peculiar behavior has only recently come to light and we
are in the process of investigating it. One avenue we are following
is to do a global continuum-chiral extrapolation to all results.
This would incorporate the constraint that fit parameters in the
chiral fits should depend smoothly on $a^2$.
Another approach is to remove the step (c) described above from our
correction procedure and leave this to be dealt with by the $a m_\ell$
extrapolation. This is possible because, at NLO in SChPT, $L_P$ does
not appear in the argument of the chiral logarithms, but only in
the prefactor.
Pending a more complete understanding, we have considered the
continuum extrapolation of results obtained after the
extrapolation to physical $a m_\ell$.

\section{Continuum Extrapolation}
\begin{figure}[t!]
\centering
\includegraphics[width=0.6\textwidth]{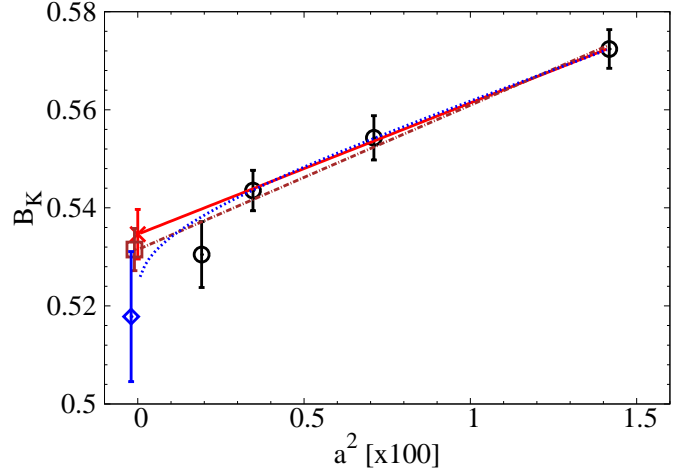}
\caption{$B_K(\mu=2\,\text{GeV})$ vs. $a^2 \, (100 \times
  \text{fm}^2)$. See text for description of data points. Note that the
  U1 point (smallest $a^2$) has not been extrapolated to physical
 $a m_\ell$. Red, brown and blue curves and points show the
  \texttt{lin3}, \texttt{lin4} and \texttt{g4-a2g2-a4} fits, respectively.
}
\label{fig:bk-su2-scale-4pt}
\end{figure}

%
%
In Fig.~\ref{fig:bk-su2-scale-4pt}, we show the results
obtained on the coarse, fine and superfine lattices after the
linear extrapolation to $a m_\ell^{\rm phys}$ plotted vs. $a^2$.
We also include the U1 result which, however, is not extrapolated
to physical $a m_\ell$, since there is only one value of
$a m_\ell$ available for the ultrafine lattices. 
Thus it cannot be consistently used in the continuum extrapolation.

Previously, when the high-statistics data from ensembles F2, F3, 
S2 and S3 were not available, we made the continuum extrapolation
using ensembles C3, F1, S1 and U1, all of which have close to
the same sea-quark masses. We then included systematic errors
accounting for the use of unphysical sea-quark masses
(the errors being 1.5\% and 1.3\% for the light and strange sea-quark
masses, respectively~\cite{swme-2}).
We found a continuum behavior which was far from linear in $a^2$
(as can be seen from Fig.~\ref{fig:bk-su2-lp:all}---the fit corresponds
to $L_P\approx 0.1\;{\rm GeV}^2$),
and we could not obtain reasonable fits including the coarse ensemble.
Dropping this, and trying various fits (along the lines described below)
we estimated an extrapolation error of 1.9\%.

Using the data extrapolated to physical $a m_\ell$ leads to a much
smoother continuum extrapolation, as can be seen in 
Fig.~\ref{fig:bk-su2-scale-4pt}. 
We can now obtain good fits including the coarse lattice result.
As noted above, we can consistently fit only to the larger three
values of $a^2$. 
Following Ref.~\cite{swme-2}, we fit both to a linear dependence
and to a five-parameter form containing dependence that we
know enters due to incomplete operator matching and higher-order
contributions.
The explicit forms are given in Table~\ref{tab:fit-func},
For the five-parameter fit we use Bayesian constraints for
$c_2-c_5$, with central values set to zero and spreads taken as
$\sigma_2=\sigma_3=\sigma_5=2 \Lambda^2$ and $\sigma_4=2$,
with $\Lambda=300\;$MeV.
These fits are shown in Fig.~\ref{fig:bk-su2-scale-4pt};
values for $\chi^2/\text{d.o.f.}$ and 
extrapolated $B_K$ are listed in the Table.
Also shown, for completeness,
is a linear fit including the U1 point, labeled {\tt lin4}.

\begin{table}[bth!]
\begin{center}
\begin{tabular}{c | c | l | c | c}
\hline
\hline
fit type & \# data & fit function & $\chi^2$/d.o.f. & $B_K(2\;{\rm GeV})$\\
\hline
{\tt lin3}       & 3 & $ c_1 + c_2 a^2 $ & 0.0307 & 0.5346(51)\\
{\tt g4-a2g2-a4} & 3 & $ c_1 + c_2 a^2 + c_3 a^2 \alpha_s(a) $  &  & \\
                 &   & $ + c_4 \alpha_s^2(a) + c_5 a^4 $ & 0.0298  & 0.5178(133)\\
{\tt lin4}       & 4 & $ c_1 + c_2 a^2 $ & 0.703  & 0.5314(43) \\
\hline
\hline
\end{tabular}
\end{center}
\caption{Details of continuum fits. For the {\tt g4-a2g2-a4} fit
with Bayesian constraints, the augmented $\chi^2$/d.o.f. is shown.
Only statistical errors are shown for $B_K$.}
\label{tab:fit-func}
\end{table}

At this stage we are not ready to quote an updated result for
$B_K$, since we clearly need to improve our understanding of
the $a m_\ell$ dependence.
Nevertheless, we note that, taking the linear continuum extrapolation
(fit {\tt lin3}) the final value for $\widehat B_K$ is $0.732$
with a $\sim 1\%$ statistical error.\footnote{%
We have also taken the opportunity to update the value
of $r_1$ to $0.3117 \, \text{fm}$, as given in Ref.~\cite{fnal-milc-1}.}
This is, in fact, consistent
with our previous result of $0.727 \pm 0.038$~\cite{swme-2}
(where the error is dominated by systematics).\footnote{%
The smallness of the shift in the central value
can be understood from Fig.~\ref{fig:bk-su2-lp:all}.
Extrapolating at physical $L_P$ there is a larger $a^2$ dependence
than at $L_P\approx 0.1\;{\rm GeV}^2$ (ignoring the coarse lattice
point in the latter case). This difference approximately
cancels the increase in the values one obtains when extrapolating
to physical $L_P$.}
Setting aside the matching error of 4.4\%, we note that the
combined $a m_\ell$ and continuum extrapolation systematics were
estimated at 2.4\%. This is certainly large enough to accommodate
any small shift we might find from our new analysis.
Thus we do not expect a significant change in our final result.
We do hope, however, that the $a m_\ell$ extrapolation error
will come down. The total error will, at present, remain dominated
by the matching error (due to our use of 1-loop perturbative matching).

In summary, results at different light sea-quark masses on
the fine and superfine MILC asqtad ensembles have uncovered 
an $a m_\ell$ dependence that is hard to understand.
An optimistic interpretation of the new results
is that the difficulties we had previously
when extrapolating at unphysical $a m_\ell$ were due to artifacts
introduced by our X-fitting and extrapolation procedure,
while the new continuum fits (shown in Fig.~\ref{fig:bk-su2-scale-4pt})
do not contain these artifacts. Analysis is underway to
test this interpretation, and final results should be available soon.


\section{Acknowledgments}
We are grateful to Claude Bernard and the MILC collaboration
for private communications.
C.~Jung is supported by the US DOE under contract DE-AC02-98CH10886.
The research of W.~Lee is supported by the Creative Research
Initiatives Program (2012-0000241) of the NRF grant funded by the
Korean government (MEST).
W.~Lee would like to acknowledge the support from KISTI supercomputing
center through the strategic support program for the supercomputing
application research [No. KSC-2011-G2-06].
The work of S.~Sharpe is supported in part by the US DOE grant
no.~DE-FG02-96ER40956.
Computations were carried out in part on QCDOC computing facilities of
the USQCD Collaboration at Brookhaven National Lab, on GPU computing
facilities at Jefferson Lab, on the DAVID GPU clusters at Seoul
National University, and on the KISTI supercomputers. The USQCD
Collaboration are funded by the Office of Science of the
U.S. Department of Energy.

\end{document}